\documentclass[conference]{IEEEtran}
\usepackage{verbatim}
\usepackage[utf8x]{inputenc}
\usepackage{epstopdf}
\usepackage[pdftex]{graphicx}
\usepackage{graphics, subcaption}
\usepackage{tabularx}
\usepackage{booktabs}
\usepackage[T1]{fontenc}
\usepackage[table]{xcolor}
\usepackage{multirow}
\usepackage{siunitx}
\usepackage{tabu}
\usepackage{booktabs}
\usepackage{ctable}
\usepackage[cmex10]{amsmath}
\usepackage{amssymb}
\usepackage{cite}
\usepackage{textcomp}
\usepackage{array}

\graphicspath{{./Figures/}}

\begin{document}
\title{Enhanced Pilot-Based Spectrum Sensing Algorithm} 

\author{\IEEEauthorblockN{Ghaith Hattab}
\IEEEauthorblockA{ECE, Queen's University\\
Kingston, ON, Canada\\
Email: g.hattab@queensu.ca}
\and
\IEEEauthorblockN{Mohammed Ibnkahla}
\IEEEauthorblockA{ECE, Queen's University\\
Kingston, ON, Canada\\
Email: mohamed.ibnkahla@queensu.ca}}

\maketitle

\begin{abstract}
In this paper, we develop an enhanced pilot-based spectrum sensing algorithm for cognitive radio. Unlike conventional pilot-based detectors which merely detect the presence of pilot signals, the proposed detector also utilizes the presence of the signal that carries the actual information. We analytically compare the performance of the proposed detector with the conventional one, and we show that the detection performance is significantly improved.
\end{abstract}


\section{Introduction}
Cognitive radio has established itself as the strongest candidate to enhance spectrum utilization by enabling secondary users (SUs) to effectively utilize the unused bands when their licensed users (also known as primary users (PUs)) are inactive \cite{Haykin, Ibnkahla2}. One of the key tasks of SUs is to reliably detect the presence or absence of the PUs using efficient detection techniques. This task is commonly known as spectrum sensing.

Spectrum sensing algorithms can be categorized into three different types: coherent detection, feature detection, and energy detection \cite{Hattab2,Yucek}. Each algorithm requires a different amount of prior information about the structure of the PU signal, and accordingly each has a different detection performance. For instance, if the SU completely knows the PU signal structure, coherent detection has the optimal performance that maximizes the output signal-to-noise ratio (SNR) \cite{Poor}. However, if the SU has absolutely no clue about the signal structure, energy detection can be implemented instead, and due to its simplicity, it has grown to be one of the most well-known spectrum sensing algorithms. Because it does not incorporate any prior information in detection, this algorithm poorly performs at the low SNR-region, a region that is a commonplace for cognitive radio applications \cite{Sahai1}. These two algorithms represent two extreme scenarios, and thus feature detectors strike a good balance between them.

In feature detection, partial information is exploited by the SU to enhance the sensing reliability (e.g., signal features). Clearly, this requires additional processing complexity compared to energy detection. Nevertheless, it is still easier to implement in comparison to coherent detection, which requires full information, and this is usually infeasible to obtain because of the existence of numerous PU networks. One of the features that can be exploited is the pilot signal. Thus, pilot-based feature detectors merely require information about the structure of this pilot signal to enhance spectrum sensing.

The conventional pilot-based sensing algorithm focuses on detecting the presence of the pilot signal \cite{Sahai1}. This is done by matching the received signal with a replica of the pilot signal, which is known \emph{a priori} at the SU side. However, this algorithm ignores the presence of the signal that carries the actual data. In general, more power is allocated to the data-carrying signal, and we show that if a detector utilizes both the presence of the pilot signal and the presence of the data-carrying signal, the detection performance can be further boosted.

In this paper, we develop a spectrum sensing algorithm, that is optimal in the Neyman-Pearson sense. The proposed detector exploits both the energy of the received signal and the presence of the pilot signal. We analytically derive the performance of the proposed detector in terms of the false alarm and miss detection probabilities. We show that the detection performance of the proposed detector outperforms the performance of the conventional one.

The rest of the paper is organized as follows. Section II presents the system model. The performance of the proposed detector is derived in Section III, and the simulation results are presented in Section IV. Finally, the conclusions are drawn in Section V.

\section{System Model}
In practical communication systems, part of the  signal power is allocated for a pilot signal to aid detection at the receiver side. For instance, pilot signals are embedded in Digital TV signals, and are set to be 11 dB weaker than the data-carrying signal \cite{ATSC}.

In spectrum sensing, the SU wants to decide if the PU is present or absent, and thus this problem is commonly described by the classical binary hypothesis testing problem, which is expressed as
\begin{equation}
\begin{aligned}
\label{eq:Model}
\mathcal{H}_0:      &   ~  y_i  =    n_i                  \\
\mathcal{H}_1:      &   ~  y_i  =    \sqrt{\theta} x_{i,p}  +   \sqrt{(1-\theta)}x_{i,d} + n_i,
\end{aligned}
\end{equation}
where $y_i$ is the received signal at the SU side, $x_{i,p}$ is the pilot signal transmitted along with the data-carrying PU signal, $x_{i,d}$, and $n_i$ is a zero-mean white Gaussian noise with variance $\sigma^2$ (i.e. $n_i\sim\mathcal{N}(0,\sigma^2)$). Here, $\theta$ denotes the fraction of the total power allocated to the pilot signal. We assume that the SU observes $N$ independent and identically distributed (IID) samples. Furthermore, we assume that $|x_{i,p}|^2=|x_{i,d}|^2=P$, where $P=(1/N)\sum_{i=1}^{N} |x_{i,p}|^2$ is the total power allocated to the PU signal (pilot and data). In other words, the power allocations to the pilot signal and the actual data are $\theta P$ and $(1-\theta)P$, respectively. We denote the SNR by $\gamma=P/\sigma^2$.

Conventionally, the following test statistic is recommended to detect the presence of the pilot signal \cite{Sahai1}
\begin{equation}
\label{eq:PilotDetector}
\Lambda_P   =   \frac{1}{N}\sum_{i=1}^{N} x_{i,p}y_i,
\end{equation}
where it is assumed that the pilot signal is orthogonal to the data-carrying signal. That is, this detector ignores the presence of the data-carrying signal.

The objective is to develop a detector that exploits the presence of the actual data as well as the prior information about the pilot signal. Because we assume no prior knowledge of the actual data, it is reasonable to assume that $x_{i,d}$ follows a zero-mean Gaussian distribution with variance $\mathbb{V}(x_{i,d})=\mathbb{E}\big[|x_{i,p}|^2\big]=P$. In other words, the data-carrying signal is modeled as $x_{i,d}\sim\mathcal{N}(0,P)$, and we assume it is independent of $n_i$. Also, because we assume that $x_{i,p}$ is known \emph{a priori}, then it is deterministic.

It can be shown that the probability distribution of the received signal under both hypotheses is
\begin{equation}
\label{eq:ModelDistribution}
\begin{aligned}
\mathcal{H}_0:  &~  y_i \sim    \mathcal{N}(0,\sigma^2)    \\
\mathcal{H}_1:  &~  y_i \sim    \mathcal{N}(\sqrt{\theta}x_{i,p},\tilde{P}+\sigma^2),
\end{aligned}
\end{equation}
where $\tilde{P}=(1-\theta)P$.

Now, we derive the optimal Neyman-Pearson (NP) detector, where the probability of miss detection is minimized with constraints on the probability of false alarm \cite{Poor}. The NP detector is the likelihood ratio, which is expressed as
\begin{equation}
\label{eq:LRT}
\prod_{i=1}^{N} \frac{f(y_i|\mathcal{H}_1)}{f(y_i|\mathcal{H}_0)},
\end{equation}
where we have used the fact that the samples are IID, and $f(y_i|\mathcal{H}_j)$ is the probability density function (PDF) of $y_i$ under hypothesis $\mathcal{H}_j$.

Using (\ref{eq:ModelDistribution}), then the PDF of one sample under $\mathcal{H}_0$ is
\begin{equation}
\label{eq:PDFH0}
f(y_i|\mathcal{H}_0)  = \frac{\exp\big(-\frac{y_i^2}{2\sigma^2}\big)}{\sqrt{2\pi\sigma^2}},
\end{equation}
and the PDF under $\mathcal{H}_1$ is
\begin{equation}
\label{eq:PDFH1}
f(y_i|\mathcal{H}_1)  = \frac{\exp\big(-\frac{(y_i-\sqrt{\theta} x_{i,p})^2}{2(\tilde{P}+\sigma^2)}\big)}{\sqrt{2\pi\big(\tilde{P}+\sigma^2\big)}}.
\end{equation}
Substituting (\ref{eq:PDFH0}) and (\ref{eq:PDFH1}) in (\ref{eq:LRT}), and then doing some straightforward simplifications, we get the following likelihood ratio test (LRT)
\begin{equation}
\frac{1}{N}\sum_{i=1}^{N}|y_i|^2
+\frac{2\sqrt{\theta}\sigma^2}{\tilde{P}}x_{i,p}y_i\overset{\mathcal{H}_1}{\underset{\mathcal{H}_0}{\gtrless}} \lambda,
\end{equation}
where $\lambda$ is the threshold that distinguishes between the two hypotheses. Therefore, the optimum Neyman-Pearson detector is expressed as
\begin{equation}
\label{eq:NPDetector}
\Lambda_{NP}=\frac{1}{N}\sum_{i=1}^{N}|y_i|^2+\frac{2\sqrt{\theta}\sigma^2}{\tilde{P}}x_{i,p}y_i.
\end{equation}
We observe that (\ref{eq:NPDetector}) is a summation of the energy of $y_i$ and the output of matching $y_i$ with the pilot signal. This is reasonable because the hypothesis testing problem in (\ref{eq:ModelDistribution}) differs in both the mean and the variance. Thus, the optimal detector exploits these two parameters to enhance the detection performance as follows. To exploit the difference in the mean, the detector correlates $y_i$ with $x_{i,p}$, and to exploit the difference in the variance, the detector computes the energy of the received signal.

Because the proposed detector exploits two parameters instead of one, we expect the detection performance to improve. Clearly, this is at the expense of  additional computational complexity. For instance, in $\Lambda_P$, we have $N+1$ multiplication operations, whereas in $\Lambda_{NP}$, we have $2N+1$ multiplication operations (excluding the multiplication of $2\sqrt{\theta}\sigma^2/\tilde P$). Nevertheless, the additional complexity is negligible thanks to the advancements in computer software and hardware.

\section{Performance Analysis}
The performance of the spectrum sensing algorithm can be analyzed in terms of the false alarm probability, $P_{FA}$, and the miss detection probability, $P_{MD}$. The former is the probability that the SU falsely decides the presence of the PU, and it is expressed as
\begin{equation}
\label{eq:Pfa}
P_{FA}  =   \mathbb{P}(\Lambda>\lambda|\mathcal{H}_0).
\end{equation}
The latter is the probability that the SU falsely decides the absence of the PU, and it is expressed as
\begin{equation}
\label{eq:Pd}
P_{MD}     =   \mathbb{P}(\Lambda<\lambda|\mathcal{H}_1).
\end{equation}
Clearly, we want to minimize $P_{FA}$ to enhance the throughput of the SU. Also, we want to minimize $P_{MD}$ to limit the harmful interference on the PUs. Unfortunately, there is an inevitable tradeoff between these probabilities, and this will be demonstrated using the complementary receiver operating characteristic (CROC), which is a plot of $P_{MD}$ versus $P_{FA}$.

\subsection{Proposed Pilot-Based Sensing Algorithm}
To find the false alarm and miss detection probabilities, the probability distribution of $\Lambda_{NP}$ must be derived. Under $\mathcal{H}_0$, we have
\begin{equation}
\Lambda_{NP}\big|\mathcal{H}_0   =   A_0 + B_0,
\end{equation}
where $A_0 =   (1/N)\sum_{i=1}^{N}|w_i|^2$, and
\begin{equation}
B_0 =   \frac{2\sqrt\theta}{N}\cdot\frac{\sigma^2}{\tilde{P}}\sum_{i=1}^{N}x_{i,p}w_i.
\end{equation}
We note that $A_0'= (N/\sigma^2)A_0$ is a central chi-square distribution with $N$ degrees of freedom (i.e. $A_0'\sim\mathcal{X}^2_{N}$). We assume that the number of observed samples is sufficiently large to approximate the chi-square distribution as a Gaussian distribution using the central limit theorem (CLT) \cite{Horgan}. Therefore, it can be shown that $A_0'\sim\mathcal{N}(N,2N)$. Similarly, let $B_0'=(N/\sigma^2)B_0$. Because $B_0'$ is a linear combination of independent Gaussian random variables, then it is still Gaussian, and it can be fully characterized by its mean and variance. We have $\mathbb{E}[B_0']=0$ since $\mathbb{E}[w_i]=0$, and the variance is
\begin{equation}
\begin{aligned}
\mathbb{V}(B_0')  &=  \frac{4\theta}{\tilde{P}^2}\sum_{i=1}^{N}\mathbb{V}(x_{i,p}w_i)\\
                    &=  \frac{4\phi\sigma^2}{\tilde{P}},
\end{aligned}
\end{equation}
where $\phi=\theta/(1-\theta)$. To summarize, we have
\begin{equation}
B_0'    \sim    \mathcal{N}\bigg(0,\frac{4\phi\sigma^2}{\tilde{P}}\bigg).
\end{equation}
Therefore, we observe that when $N$ is large, $\Lambda_{NP}\big|\mathcal{H}_0$ can be approximated as a Gaussian random variable. That is, we have
\begin{equation}
(N/\sigma^2)\Lambda_{NP}\big|\mathcal{H}_0 \sim\mathcal{N}\Big(N,\frac{2N}{\tilde{P}}\big[\tilde{P}+2\phi\sigma^2\big]\Big).
\end{equation}

Following the same procedure, we have under $\mathcal{H}_1$
\begin{equation}
\Lambda_{NP}\big|\mathcal{H}_1   =   A_1 + B_1,
\end{equation}
where $A_1=(1/N)\sum_{i=1}^{N}|y_i|^2$ and
\begin{equation}
B_1 =   \frac{2\sqrt\theta}{N}\cdot\frac{\sigma^2}{\tilde{P}}\sum_{i=1}^{N}x_{i,p}y_i.
\end{equation}
Note that $A_1'= [N/(\tilde{P}+\sigma^2)]A_1$ is a non-central chi-square distribution with $N$ degrees of freedom and a non-centrality parameter of $\eta$, which is expressed as
\begin{equation}
\eta =    \frac{\theta NP}{\tilde{P}+\sigma^2}.
\end{equation}
That is, $A_1'\sim\mathcal{X}^2_{N}(\eta)$. Using the CLT, we can approximate this random variable as a Gaussian one, where  $A_1'\sim\mathcal{N}\big(N+\eta,2N+4\eta\big)$. We can expand this such that
\begin{equation}
A_1'\sim\mathcal{N}\bigg(\frac{N(P+\sigma^2)}{\tilde{P}+\sigma^2},
\frac{2N[(1+\theta)P+\sigma^2]}{\tilde{P}+\sigma^2}\bigg).
\end{equation}
Similarly, let $B_1'=[N/(\tilde{P}+\sigma^2)]B_1$, then $B_1'$ has a Gaussian distribution with mean
\begin{equation}
\mathbb{E}[B_1']        =   \frac{2N\phi\sigma^2}{\tilde{P}+\sigma^2}.
\end{equation}
Furthermore, the variance is expressed as
\begin{equation}
\begin{aligned}
\mathbb{V}(B_1')  &=  \frac{4\theta\sigma^4}{\tilde{P}^2(\tilde{P}+\sigma^2)^2}\sum_{i=1}^{N}\mathbb{V}(x_{i,p}y_i)   \\
                  &=  \frac{4N\phi\sigma^4}{\tilde{P}(\tilde{P}+\sigma^2)}.
\end{aligned}
\end{equation}
To summarize, we have
\begin{equation}
B_1'    \sim    \mathcal{N}\bigg(\frac{2N\phi\sigma^2}{\tilde{P}+\sigma^2}, \frac{4N\phi\sigma^4}{\tilde{P}(\tilde{P}+\sigma^2)}\bigg).
\end{equation}
When $N$ is large, $\Lambda_{NP}\big|\mathcal{H}_1$ can be approximated as a Gaussian random variable. It can be shown that
\begin{equation}
[N/(\tilde{P}+\sigma^2)]\Lambda_{NP}\big|\mathcal{H}_1 \sim\mathcal{N}(m,v),
\end{equation}
where
\begin{equation}
m   =   \frac{NP+N(1+2\phi)\sigma^2}{\tilde{P}+\sigma^2},
\end{equation}
and
\begin{equation}
v  = \frac{2N\big[(1-\theta^2)P^2+(\tilde{P}+2\phi\sigma^2)\sigma^2\big]}{\tilde{P}(\tilde{P}+\sigma^2)}.
\end{equation}

Finally, the probability of false alarm is expressed as
\begin{equation}
\label{eq:NPPfa}
P_{FA}^{NP} =   Q\Bigg(\frac{\sqrt{N/2}(\lambda_{NP}-1)}{\sqrt{1+2\phi\tilde\gamma^{-1}}}\Bigg),
\end{equation}
where $\tilde\gamma=(1-\theta)\gamma$, and $Q(.)$ is the tail probability of the cumulative density function (CDF) of the standard Gaussian. Similarly, the probability of miss detection is expressed as
\begin{equation}
\label{eq:NPPMD1}
P_{MD}^{NP} =   1-Q\Bigg(\frac{\sqrt{N/2}[\lambda_{NP}-\gamma-(1+2\phi)]}
{\sqrt{(1+\tilde\gamma^{-1})[(1-\theta^2)\gamma^2+\tilde\gamma+2\theta]}}\Bigg).
\end{equation}
Note that from (\ref{eq:NPPfa}), we have $\sqrt{N/2}(\lambda_{NP}-1)=\sqrt{1+2\phi\tilde\gamma^{-1}}Q^{-1}\big(P_{FA}^{NP}\big)$. Thus, we can rewrite (\ref{eq:NPPMD1}) as
\begin{equation}
\label{eq:NPPMD2}
P_{MD}^{NP} =   Q\Bigg(\frac{\sqrt{N/2}(\gamma+2\phi)-\sqrt{1+2\phi\tilde\gamma^{-1}}Q^{-1}\big(P_{FA}^{NP}\big)}
{\sqrt{(1+\tilde\gamma^{-1})[(1-\theta^2)\gamma^2+\tilde\gamma+2\theta]}}\Bigg),
\end{equation}
where we have used $Q(x)=1-Q(-x)$.

\subsection{Conventional Pilot-Based Sensing Algorithm}
Ignoring the data-carrying signal, it can be shown that
\begin{equation}
\label{eq:PilotPDF}
\begin{aligned}
\mathcal{H}_0:  &~  \Lambda_P \sim    \mathcal{N}(0,(P\sigma^2)/N)    \\
\mathcal{H}_1:  &~  \Lambda_P \sim    \mathcal{N}(\sqrt{\theta}P,(P\sigma^2)/N).
\end{aligned}
\end{equation}
Thus, the probability of false alarm is
\begin{equation}
\label{eq:PilotPfa}
P_{FA}^P  =   Q\bigg(\frac{\lambda_P}{\sqrt{(P\sigma^2)/N}}\bigg),
\end{equation}
and the probability of miss detection is
\begin{equation}
\label{eq:PilotPd1}
P_{MD}^P     =   1-Q\bigg(\frac{\lambda_P-\sqrt{\theta} P}{\sqrt{(P\sigma^2)/N}}\bigg).
\end{equation}
By eliminating $\lambda_P$ from (\ref{eq:PilotPfa}) and (\ref{eq:PilotPd1}), we have
\begin{equation}
\label{eq:PilotPd2}
P_{MD}^P   =  Q\bigg(\sqrt{N\theta\gamma}-Q^{-1}(P_{FA}^P)\bigg)
\end{equation}

\section{Simulation Results}
In this section, we analyze the performance of both the proposed detector and the conventional one in terms of the false alarm and miss detection probabilities. We assume here that $\theta=0.1$.

\begin{figure}[!b]
\centering
\includegraphics[width=3.45in]{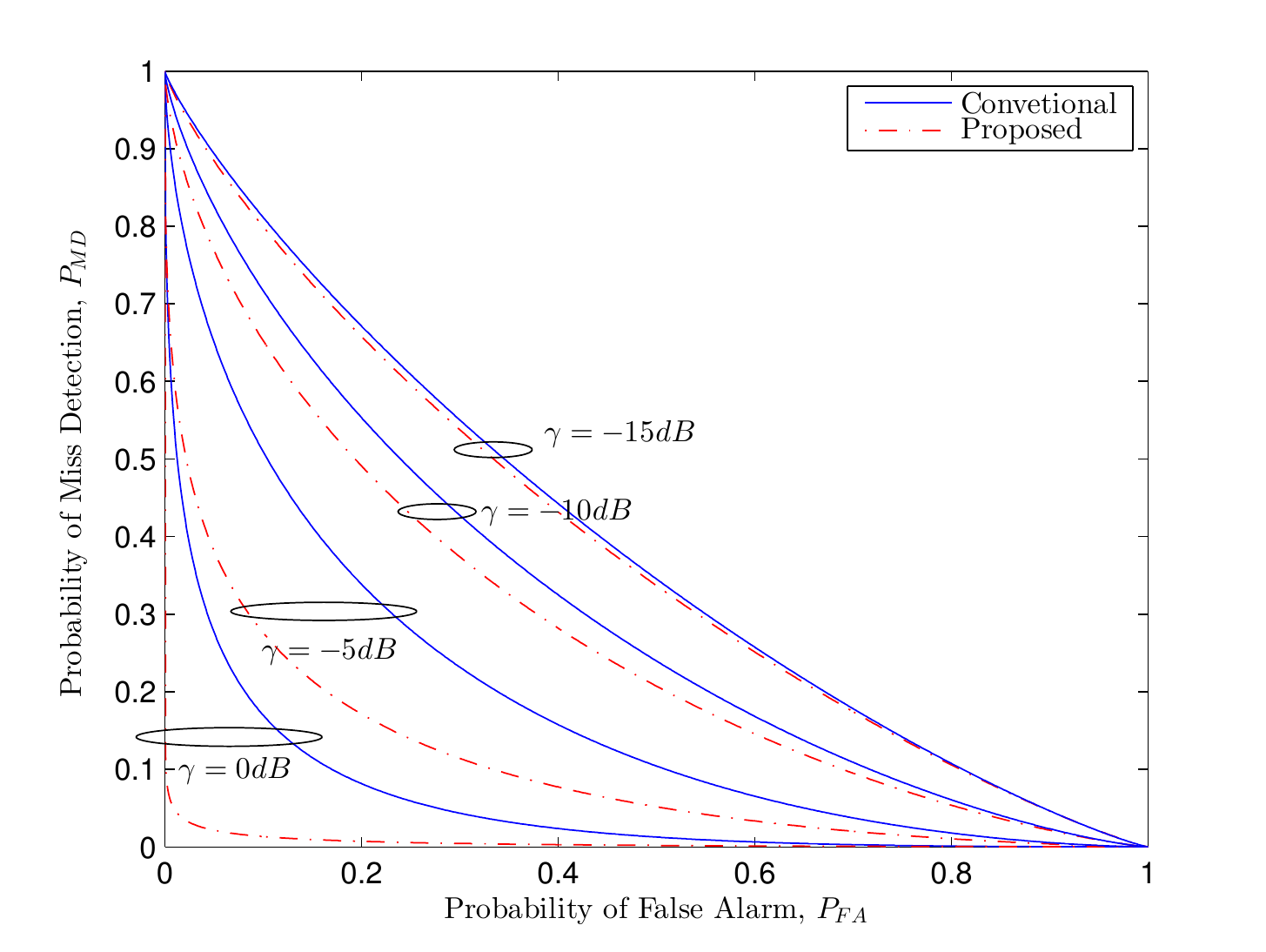}
\caption{CROC curves for different SNR levels. }
\label{fig:CROC}
\end{figure}

Fig. \ref{fig:CROC} illustrates the CROC curves for different SNR values. We assume that the SU observes $N=100$ samples. We have the following remarks. It is observed that at moderately-low SNR-region (e.g., $\gamma=-5$ dB), the proposed detector significantly outperforms the conventional detector. The gains diminish as the SNR decreases, and in particular, at the very low SNR-region (e.g., $\gamma=-15$ dB), the performance of the proposed detector slightly outperforms the conventional one. This can be explained as follows. Recall that the proposed detector consists of two components: an energy-based component and a feature-based one. At the very low SNR-region, the energy-based component has a very poor performance, and hence its contribution becomes negligible. In other words, the proposed detector does not gain from it, and hence it merely relies on the feature-based component. Consequently, the performance of the proposed detector is very close to the conventional one at the very low SNR-region.

\begin{figure}[!t]
\centering
\includegraphics[width=3.5in]{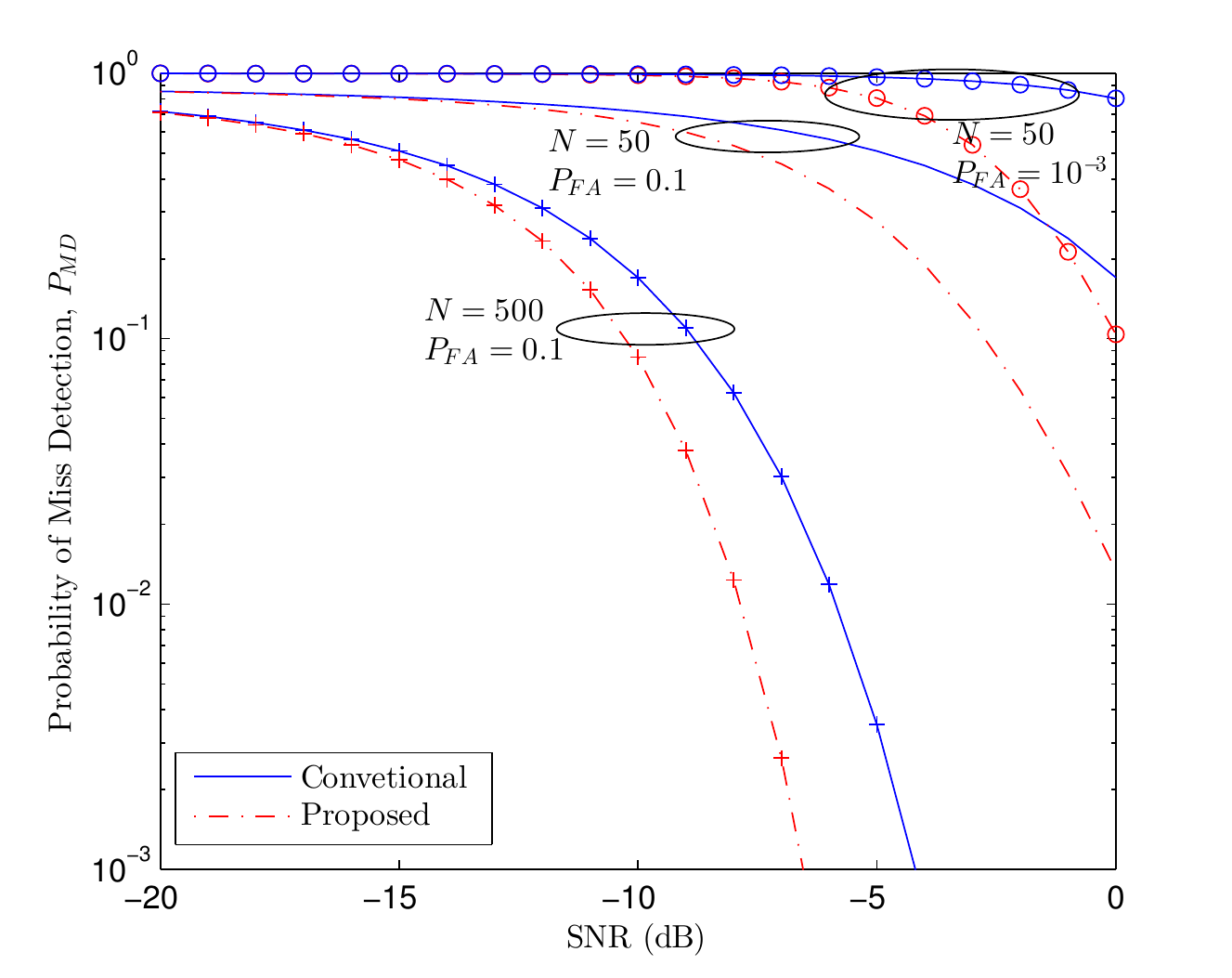}
\caption{$P_{MD}$ versus SNR under different number of samples and false alarm requirements.}
\label{fig:MissProbability}
\end{figure}

Fig. \ref{fig:MissProbability} illustrates the performance of the probability of miss detection with variations of SNR. We observe the following. Clearly, the miss detection probability is less for the proposed detector compared to the conventional one. Also, we observe that the proposed detector is more robust when we need a tighter false alarm probability (e.g., $P_{FA}=0.001$). Finally, increasing the number of samples significantly reduces the miss detection probability, but such parameter must be carefully tuned because of the fundamental tradeoff between sensing time and throughput (i.e. increasing $N$ improves the detection at the expense of the SU's throughput)  \cite{Liang}. Alternatively, the proposed detector requires fewer samples, compared to the conventional one, to achieve a predetermined $P_{FA}$ and $P_{MD}$.

\section{Conclusion}
Pilot-based detectors strike a good balance between energy detection and coherent detection. In particular, it requires partial information (instead of full information), and it is more robust at low SNR (compared to energy detection). In this paper, we propose an enhanced pilot-based spectrum sensing algorithm. Unlike conventional detectors, the proposed detector does not only search for the pilot signal, but it also computes the energy of the entire signal.  We show that the probability of miss detection can be significantly reduced especially for tight false alarm requirements. These results motivate sharing information about pilot signals between the PU and the SU networks in order to maintain reliable detection, which enhances the throughput for the SU and limit the interference on the PUs.

\bibliographystyle{IEEEtran}
\bibliography{C:/Users/Ghaith_90/Dropbox/Latex/IEEEabrv,C:/Users/Ghaith_90/Dropbox/Latex/References}
\end{document}